\documentclass[a4paper,12pt]{article}

\usepackage[utf8]{inputenc}
\usepackage[english]{babel}
\usepackage[T1]{fontenc}
\usepackage[dvips,final]{graphicx}
\usepackage{float}
\usepackage{amsmath}
\usepackage{natbib}

\newcommand{\dd}{\mathrm{d}}

\newcommand{\bel}[1]{\begin{equation} \label{#1}}
\newcommand{\ee}{\end{equation}}

\newcommand{\matlab}{\textsc{Matlab}}

\begin{document}

\thispagestyle{empty}
\title{Current source density reconstruction \\ 
from incomplete data}

\author{Daniel K. Wójcik, Szymon Łęski \\
\\ {}
\\ Department of Neurophysiology, 
\\ Nencki Institute of Experimental Biology, 
\\ 3 Pasteur Street, 02-093 Warsaw, Poland}

\maketitle

\section*{}

{\bf We propose two ways of estimating the current source density
  (CSD) 
  from measurements of voltage on a Cartesian grid with missing
  recording points using the inverse CSD method. The 
  simplest approach is to substitute local averages (LA) in place of
  missing data. 
  A more elaborate alternative is to estimate a smaller number of CSD
  parameters than the actual number of recordings and to take the
  least-squares fit (LS). We compare the two approaches in the three
  dimensional case on several sets of surrogate 
  and experimental
  data, for varying
  numbers of missing data points, and discuss their advantages and
  drawbacks.  
  One can construct CSD distributions for which one or the other approach is better. However, 
  in general, LA method is to be recommended being more stable and more robust to variations in the recorded fields.
}

\section{Introduction}
\label{sec:introduction}

A common measure of neural population activity is the local field
potential (LFP), the low-frequency part of the extracellular electric
potential~\citep{Nunez2005}.  The LFP is generated by trans-membrane
currents in neighboring cells which are usually described on a
coarse-grained level by the current source density
(CSD)~\citep{Plonsey1969, Nicholson1971, Mitzdorf1985, Nunez2005}. In
the quasi-static approximation the relation of the CSD, $C$, to the
potentials, $\phi$, is
\begin{equation}
\label{lapl}
\nabla (\sigma\nabla\phi) = -C,
\end{equation}
where $\sigma$ is the electrical conductivity tensor, which for
simplicity we assume to be a constant scalar (isotropic, homogeneous
medium). One consequence of this equation is non-locality: $\phi$ is
not trivial even in regions where $C=0$. This means that the recorded
LFP may reflect the activity of quite distant cells.

When the recordings of LFP at several locations are available one can
attempt reconstruction of the CSD which generated
them~\citep{Mitzdorf1985}. Such recordings can be obtained e.g. with
an electrode with multiple contacts or with a two-dimensional
multi-electrode array~\citep{Csicsvari2003,Bartho2004,Buzsaki2004}.

The simplest method to calculate CSD is to use a numerical
approximation of the second derivative of the
potential~\citep{Nicholson1971, Mitzdorf1985}, e.g. in case of 1-D
electrode with equidistant contact points spaced by $h$ one obtains
(for interior contacts):
\begin{equation}
\label{tradcsd}
C(z_i) = -\sigma \frac{\phi(z_i+h)-2\phi(z_i)+\phi(z_i-h)}{h^2}.
\end{equation}
Such an approach has several disadvantages. One of them is that
Eq.~(\ref{tradcsd}) cannot be applied to boundary points. This is
particularly inconvenient in case of two- or three-dimensional data,
where the boundary may comprise majority of the
points~\citep{Leski2007}.

Another method for estimating the CSD is the inverse CSD (iCSD)
method~\citep{Pettersen2006, Leski2007}. Here one does not try to use
Eq.~(\ref{lapl}) directly (which is the case in traditional CSD). 
Instead, the idea is to establish a one-to-one relation $F$ between
measured voltages and CSD distributions \emph{via} inversion of the
forward solution. This is achieved in the following way: assume $N$
recording points on a Cartesian grid (one-, two- or
three-dimensional). Consider $N$-parameter family of CSD distributions
--- this means that given the values of the $N$ parameters one can
assign a value of CSD to each spatial position.  Then the values of
the potential, $\phi$, on the grid can be obtained by solving a
well-posed boundary value problem related to the elliptic partial
differential equation, Eq.~(\ref{lapl}) (forward solution). Therefore,
the $N$ measured voltages are functions of the CSD parameters. If the
family and the parameterization are chosen well, one can invert this
relation and from the $N$ measured potentials recover the $N$
parameters of CSD. Usually one parameterizes the CSD with its values
on the measurement grid and interpolates between the grid points,
linearly or with splines, but there are more
possibilities~\citep{Pettersen2006, Leski2007}.

In the reconstruction one may assume that the CSD is non-zero only
inside the grid. Usually, however, the actual CSD 
extends in the tissue beyond the grid set by the measurement points
and such an assumption would lead to large reconstruction
errors~\citep{Leski2007}. To avoid them one may use a trick of
extending the grid spanning the CSD by an additional layer.  One can
set the CSD values at the additional nodes to zero or duplicate the
value from the neighboring node of the original
grid. In~\citep{Leski2007} these two approaches were denoted by B or D
boundary conditions, respectively, and it was shown that they improve
the reconstruction quality. Note that the new CSD family is still
parameterized with its $N$ values at the original grid and that the
intention of such a procedure is to improve the reconstruction
fidelity inside the grid and not to estimate the CSD outside the grid.

\section{Inverse CSD on incomplete data}
\label{sec:incomp}

A practical problem in the application of iCSD method to real datasets
is how to deal with missing recording points.  Such cases arise
surprisingly often in real experiments. There may be several reasons
for this: a contact of a multi-electrode may not be functioning, some
channels may be used for other purposes (e.g. stimulation), or the
experiment may be terminated early before all the data are collected.

One way to deal with such data is to patch them with the mean of the
neighboring potentials of a missing contact~\citep{Leski2008}.  We
denote this method by LA for local averages.  This approach means
replacing the missing true potential at a point with a linear
approximation estimated from the neighbors.

A more elaborate alternative is to reduce the size of the CSD grid and
find the least-squares solution to such an overdetermined system. That means 
choosing such values of the parameters of the CSD spanned on a
smaller grid which minimize the sum of squared differences in
potential at all the available electrode points. We denote this method
by LS for least squares. The advantage of this approach is that we use
only the available data without making any assumptions about the
missing recordings, so it seems to be better motivated than
LA. However, we decrease the spatial resolution of the reconstructed
CSD so it is hard to tell \emph{a priori} which method is better.

\section{Gaussian sources}
\label{sec:results}

\begin{figure}[htbp]
  \begin{center}
    \hspace{5em} A)
    \hspace{-5em}\includegraphics[width=0.8\textwidth]{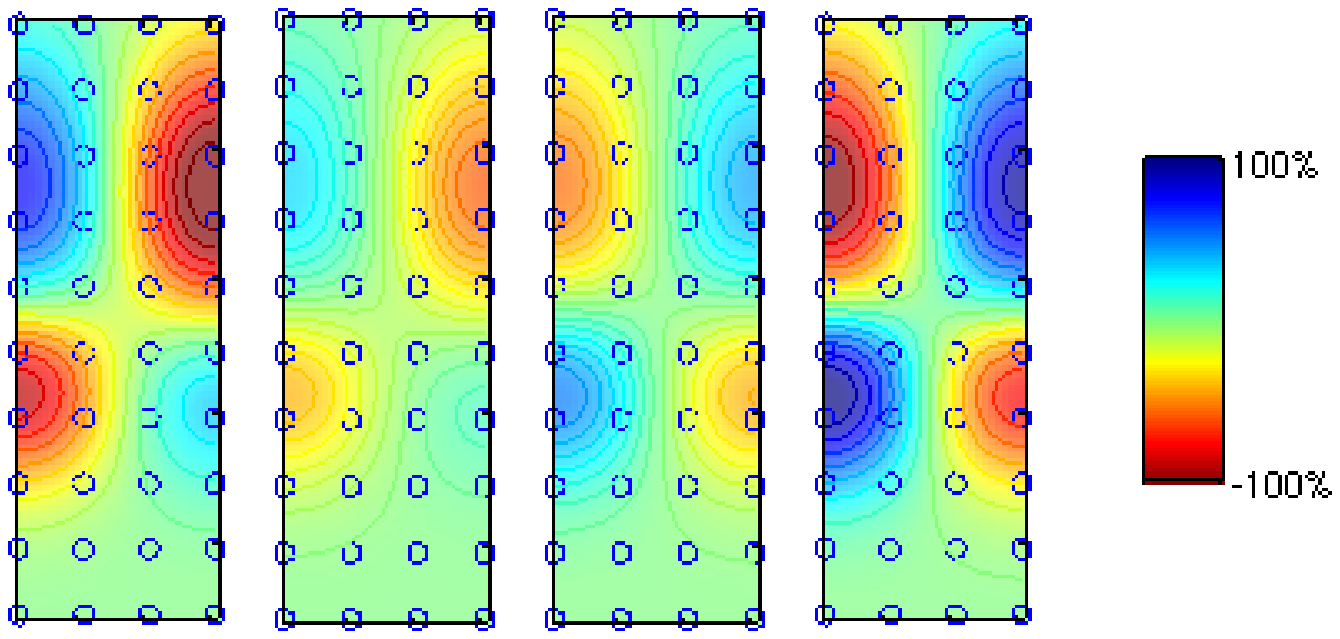}\\ 
    B)\includegraphics[width=0.4\textwidth]{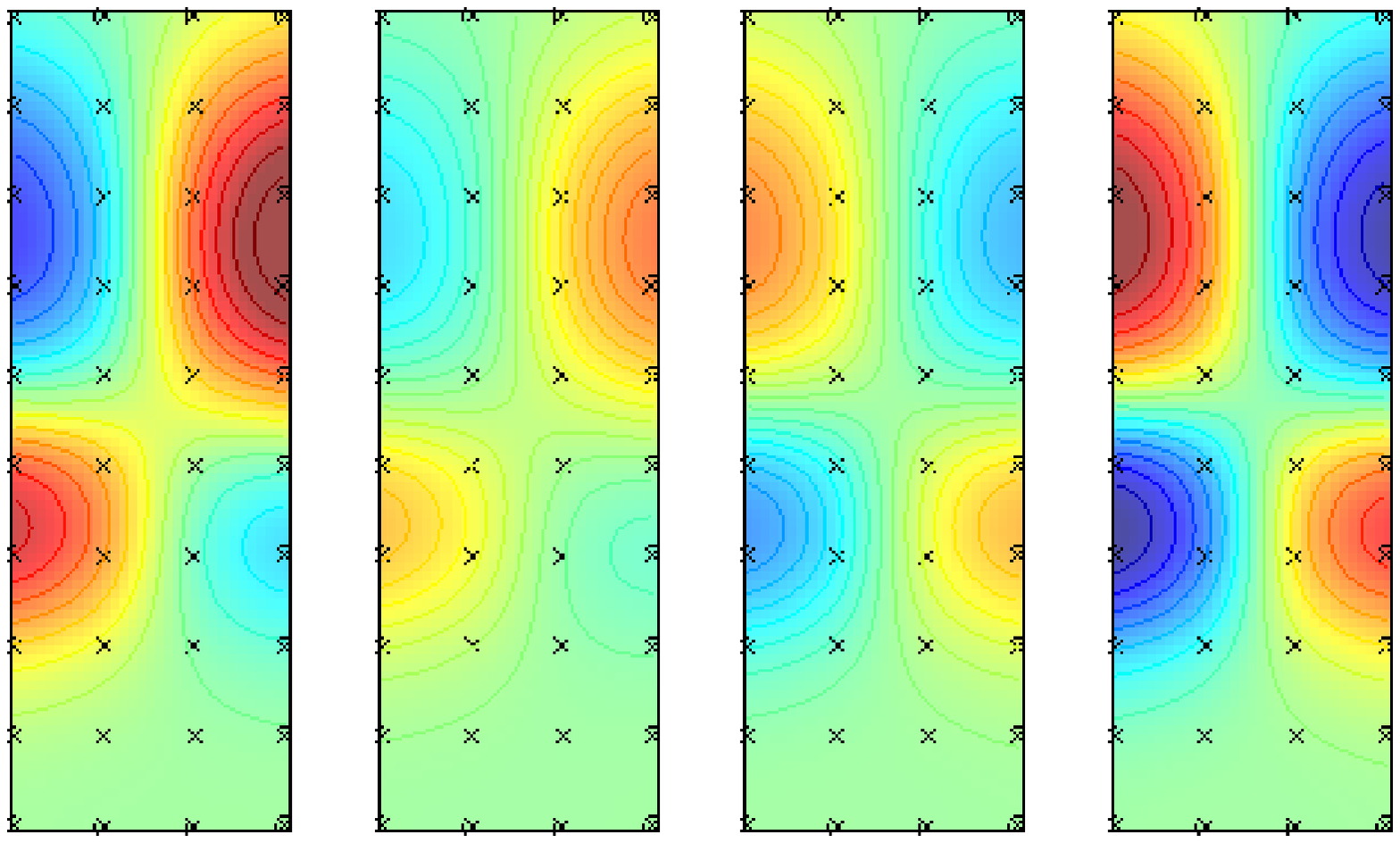} \hspace{1em}
    C)\includegraphics[width=0.4\textwidth]{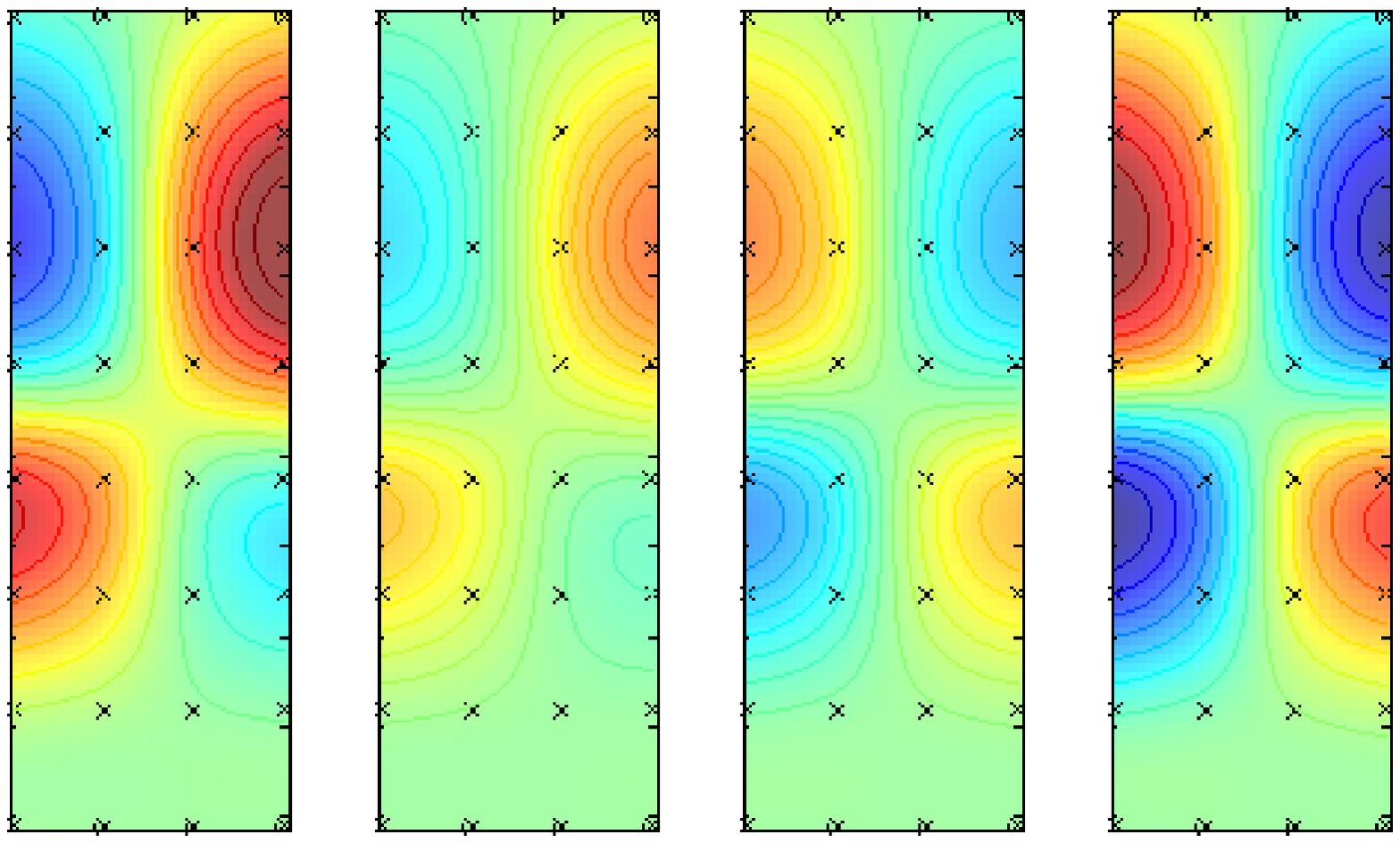}
  \end{center}

  \caption{\label{figg0}A) Gaussian sources studied in Section
    \ref{sec:results}, four consecutive slices ($x=1\ldots4$) through
    the volume. Electrode positions are marked with circles. B)
    Reconstruction of the CSD distribution spanned on the full $4
    \times 10 \times 4$ grid from the set of potentials calculated at
    the nodes of the grid (denoted by x's).  C) Reconstruction using
    LS method on complete data; spanned on a smaller, $4 \times 8
    \times 4$ grid.}
\end{figure}

To find out which of the proposed approaches works better we first
tested them both on three-dimensional surrogate Gaussian sources
(Fig.~\ref{figg0}).
We calculated the potentials on a grid of $(x,y,z) \in 4 \times 10
\times 4$ equally spaced points. Details of the structure of the
sources and calculation of LFP are given in the Appendix. This choice
of the sources and the grid was motivated by a recent experimental
study of evoked potentials in the barrel cortex of the
rat\footnote{J. Kami\'nski, private communication. Note that there are
  more sophisticated models of the cortical LFP,
  e.g.~\citep{Tenke1993}, but for these tests we found our simple model
  adequate.}. The sources were elongated along the $y$ axes so that
the conclusions would hold for 
cortical dipoles generated by active pyramidal cells.

For the tests we removed a number of virtual `recording points',
reconstructed the CSD using both LA and LS methods, and compared the
normalized $L^2$ reconstruction errors~\citep{Leski2007}: \( e = \int
(C-\hat{C})^2 \dd x/ \int C^2 \dd x, \) where $C$ is the original and
$\hat{C}$ is the reconstructed CSD. For the LS method we used a grid
of $4 \times 8 \times 4$ points which covered the whole space occupied
by the original grid. This implied larger spacing in the $y$
direction. The iCSD reconstruction was performed with the
\matlab\mbox{ } scripts from \cite{Leski2007}, modified for the
situation at hand. We used not-a-knot splines with D boundary
conditions.\footnote{By D boundary conditions we mean solution on a
  larger grid with one extra layer added in every direction beyond the
  original. We assume identical CSD at the added layer and its nearest
  neighbor in the original grid. 'Not-a-knot' splines are the cubic
  splines implemented in \matlab, they differ from `normal' splines in
  the conditions at the extreme points. See~\cite{Leski2007} for
  details.}
\begin{figure}[htbp]
  \begin{center}
    A) \includegraphics[width=0.3\textwidth]{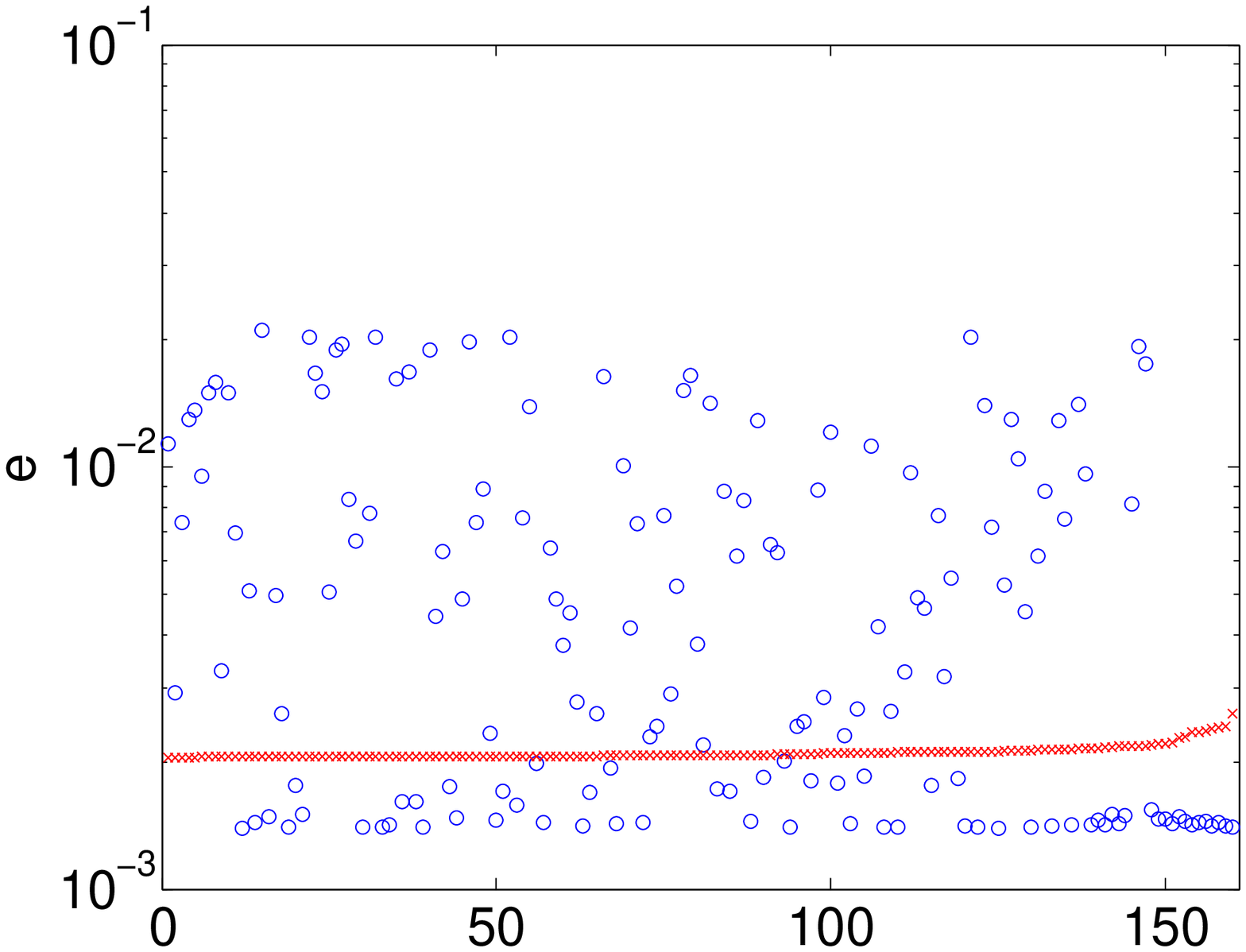}
    B) \includegraphics[width=0.3\textwidth]{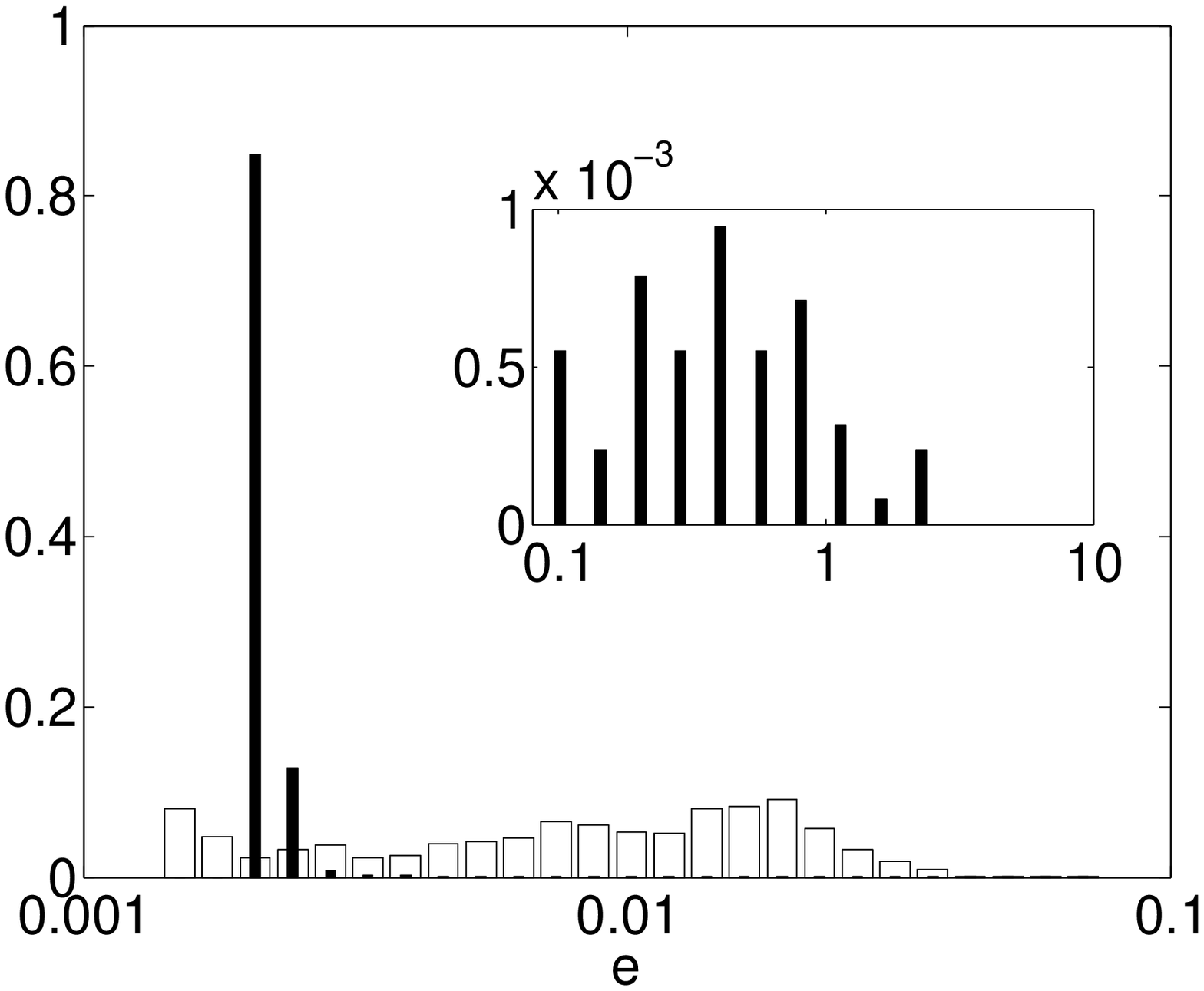}\\
    C) \includegraphics[width=0.3\textwidth]{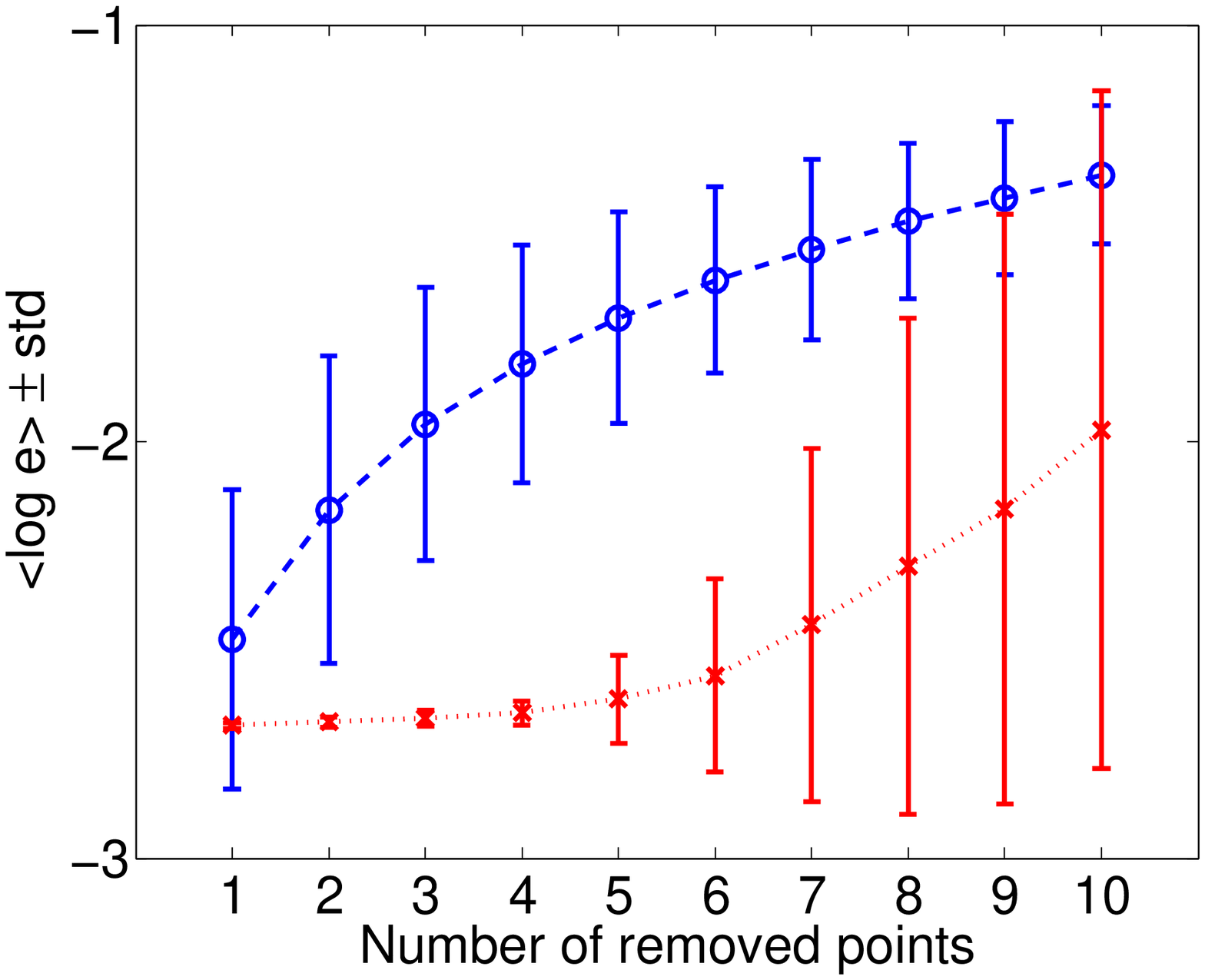}
    D) \includegraphics[width=0.3\textwidth]{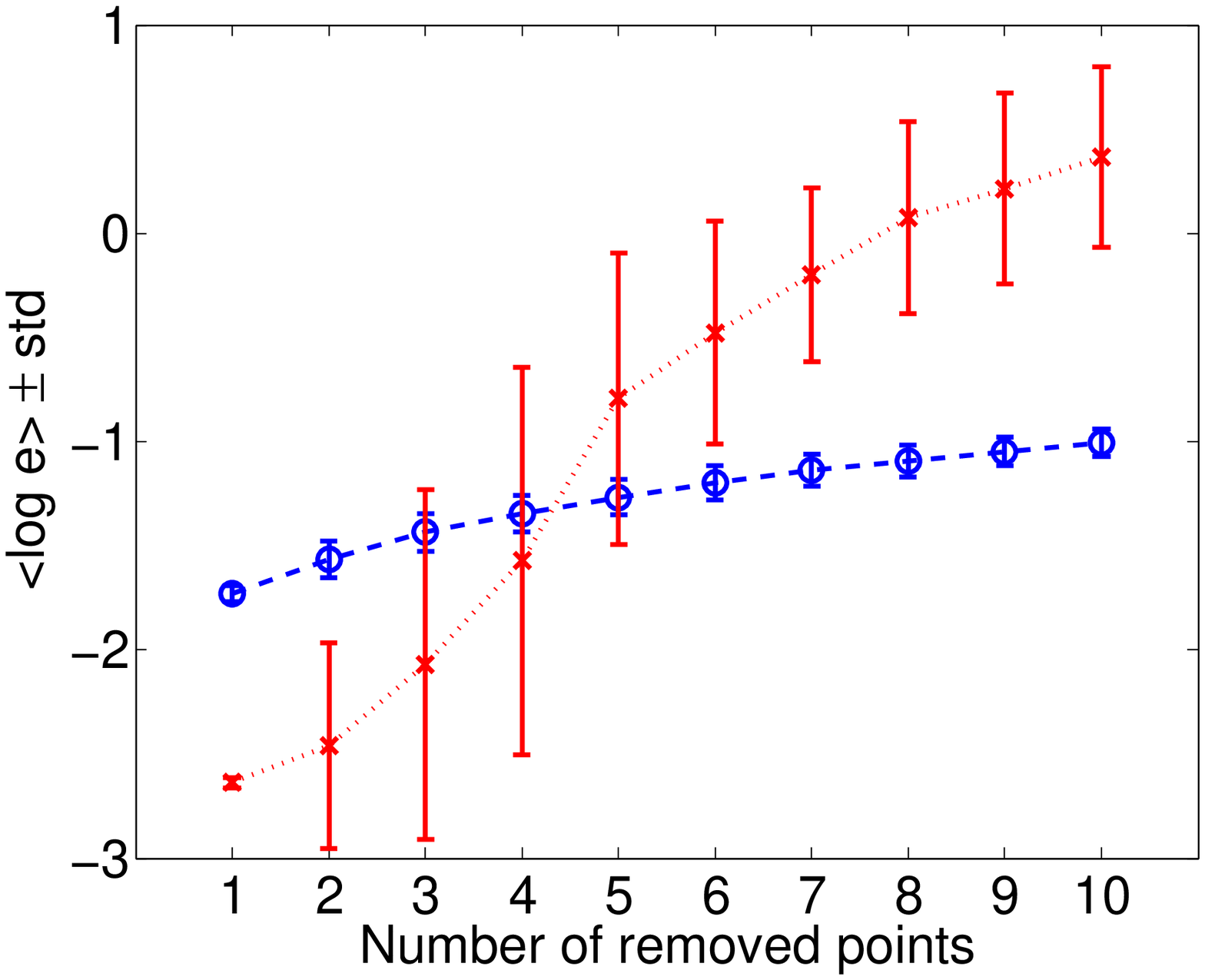}
  \end{center}

  \caption{\label{figg}Comparison of local averages (LA) and least
    squares (LS) methods of reconstructing CSD from incomplete
    data. A) Results of reconstruction in the case of a single
    recording point removed from the grid. Normalized reconstruction
    error for all 160 possibilities, LS: x's, LA: o's, sorted
    according to the LS error. B) Histogram of normalized
    reconstruction errors for a pair of grid points removed. Thick
    bars: LA, thin bars: LS. Inset: outliers in the LS method. C)
    Comparison of LA (o's) and LS (x's) methods for varying number $n$
    of recording points removed from the grid (X axis). Y axis:
    average logarithm of normalized reconstruction error, error bars
    are $\pm$ standard deviation, for the best 90\% out of 2000 random
    choices of removed points (except $n=1$ where 90\% of all 160
    possibilities are taken). D) Same as C, but for the worst 10\% of
    the cases.}
\end{figure}

Standard iCSD reconstruction from the calculated potentials 
gives reconstruction error $e$ of 0.14\%. This indicates the quality
of the reconstructed approximation of the smooth Gaussian sources by a
set of splines spanned on the grid of recorded points. The LS method
applied for the reduced grid gives $e=0.21\%$ which shows how little
information is lost when the number of nodes of the grid used for the
reconstruction is reduced by 20\%, see Figure~\ref{figg0}. This is
possible in this case thanks to the relatively large extent and slow
variation of the sources in the $y$ direction: the sparser grid is
still dense enough to effectively sample the sources. In general, the
degradation of the reconstruction quality caused by using a sparser
grid will strongly depend on how rapidly the CSD varies in space, see
Section~\ref{sec:expdata}.

We have scanned all the 160 cases of one recording point
withdrawn. The LS method gives stable reconstruction error from
$e=0.21\%$ to $e=0.26\%$. The error of the LA method ranges from
$e=0.14\%$ (which means that the missing datum was indeed the mean of
its neighbors) up to 2.1\%, see Fig.~\ref{figg}A.

There are 12720 possible choices of a pair of electrodes for the set
of recording points considered. Here the results are more intricate:
as before, LS typically gives smaller errors (Fig.~\ref{figg}B), but
from time to time a huge error occurs, with $e$ reaching 270\% (63
outliers shown in an inset in Fig.~\ref{figg}B). The outliers can
occur only for specific configuration of the missing pair
$(x_1,y_1,z_1)$, $(x_2,y_2,z_2)$. The necessary (but not sufficient)
conditions for large errors are $(x_1,z_1)=(x_2,z_2)$ and $(y_1,
y_2)\in \{(1,2), (1,3), (2,3), (9,10), (8,10), (8,9)\}$. There are
only 96 of such troublesome pairs and all the outliers in
Fig.~\ref{figg}B are of this type.

With growing number of missing recording points the reconstructions
become less and less reliable with the LS method becoming
monotonically worse with respect to the LA method, Fig.~\ref{figg}C,
D. Interestingly, the distributions of the results qualitatively have
the same character as in the case of two missing electrodes: that is,
the errors of LA method have a unimodal distribution while the
distributions of errors in the LS approach have two modes, one with
results better than for LA, the other with extremely large errors. The
mean error of the LA method also grows but huge errors do not
occur. Figures~\ref{figg}C, D show the results obtained for 2000
random choices of the missing recording points.

\section{Experimental data}
\label{sec:expdata}

The second part of the test of the two methods was performed on
three-dimensional recordings in the rat forebrain of potentials evoked
by the deflection of a bunch of whiskers~\citep{Leski2007}. The
recordings were made on a grid of $4\times5\times7$ points. Here we
analyse the same two representative latencies which were used as
illustrations in~\cite{Leski2007}, where the dataset is described in
detail.

We perform the same analysis as for the Gaussian sources, apart from
the fact that now we do not know the real CSD. Therefore, as the
reference $C$ we take the reconstruction spanned on the full
$4\times5\times7$ grid calculated from the full set of
recordings. Such $C$ is the best representation of real CSD in the
tissue available to us.

\begin{figure}[htbp]
  \begin{center}
    \begin{tabular}[b]{c}
      A)\\[12.5em] B)
    \end{tabular}
    \includegraphics[width=0.8\textwidth]{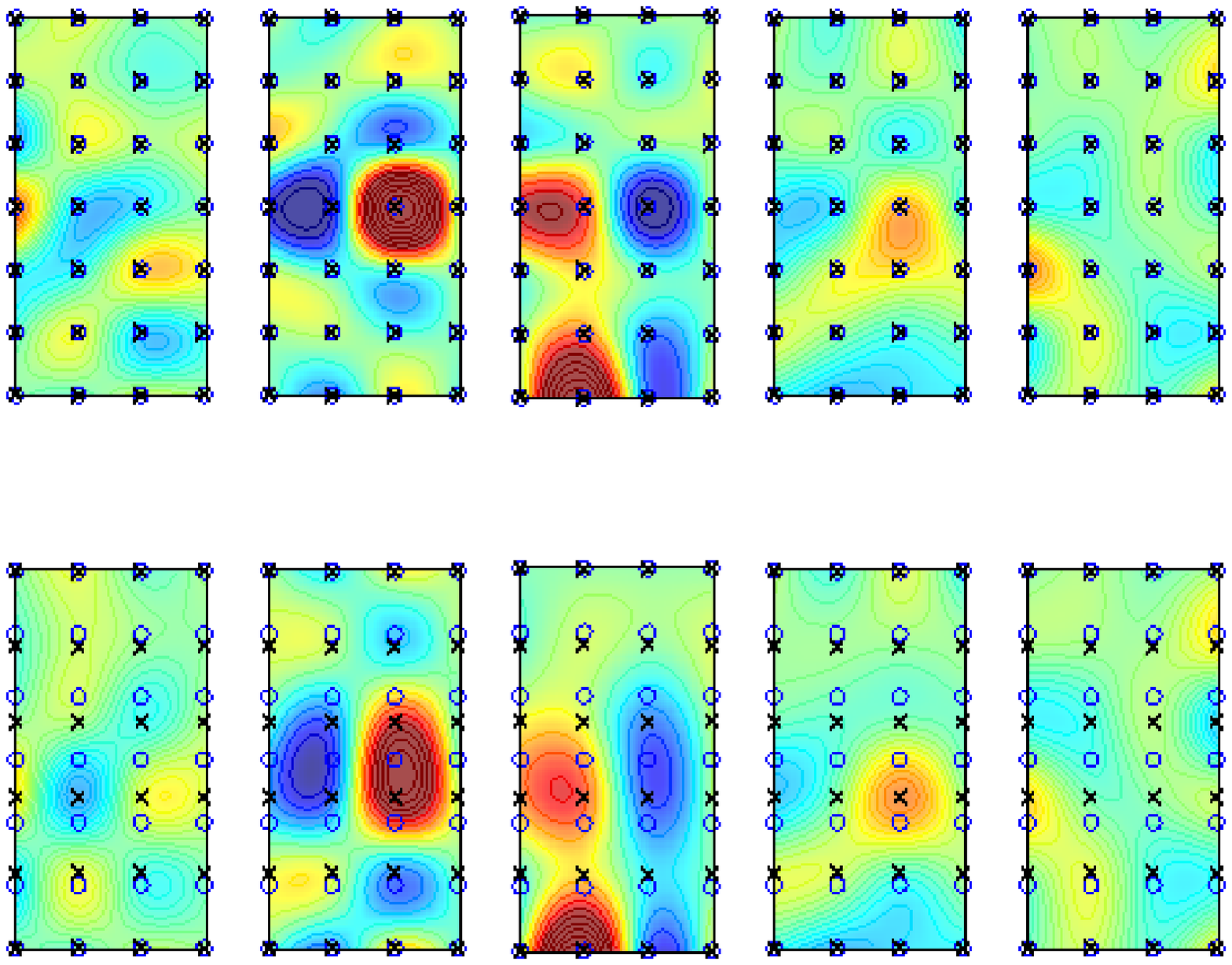}
\mbox{ } \raisebox{4em}[0pt]{
\includegraphics[width=0.1\textwidth]{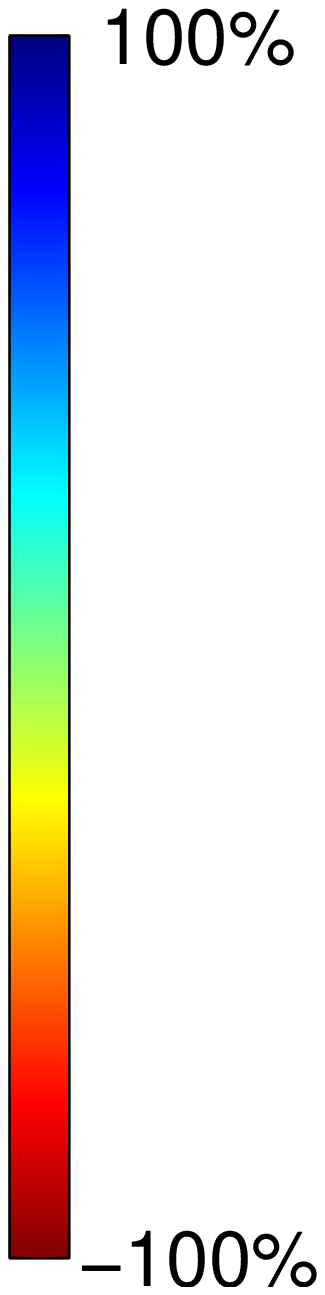}}
  \end{center}
  \caption{\label{figg20}Reconstructions of CSD from experimental
    data, $t=3.5$ms after stimulation of vibrissa. Each row presents a
    three-dimensional region of the rat forebrain. The electrode
    positions ($4\times5\times7$ grid) are marked with circles, nodes
    of the grid are marked with x's. A) The reference data set: CSD
    reconstructed on the full electrode grid. B) CSD reconstructed
    from the complete set of recordings but spanned on a sparser
    $4\times5\times6$ grid. Note that some sources are not adequately
    sampled using the sparser grid. }
\end{figure}

Figure~\ref{figg20}A shows the reference data set and
Fig.~\ref{figg20}B shows the CSD reconstructed by LS method on a
sparser, $4\times5\times6$ grid, from the complete set of recorded
potentials, $3.5$ms after the stimulus onset. Already here we can
observe how the intricate structure of activation in the tissue is
distorted ($e=21\%$) when using a smaller spanning grid which was not
the case for the Gaussian sources modeling the cortical CSD
(Sec.~\ref{sec:results}).

\begin{figure}[htbp]
  \begin{center}
    A) \includegraphics[width=0.3\textwidth]{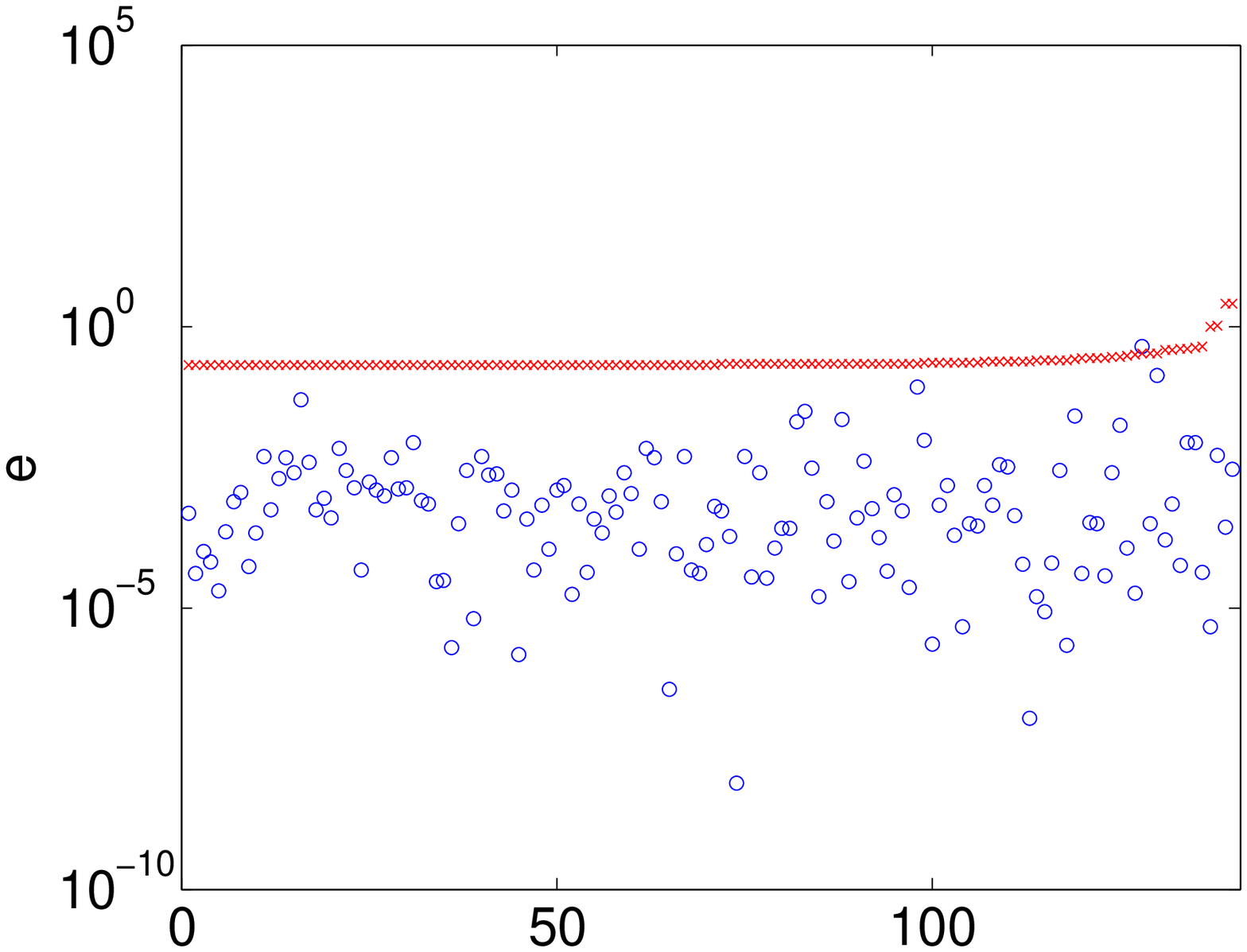}
    B) \includegraphics[width=0.3\textwidth]{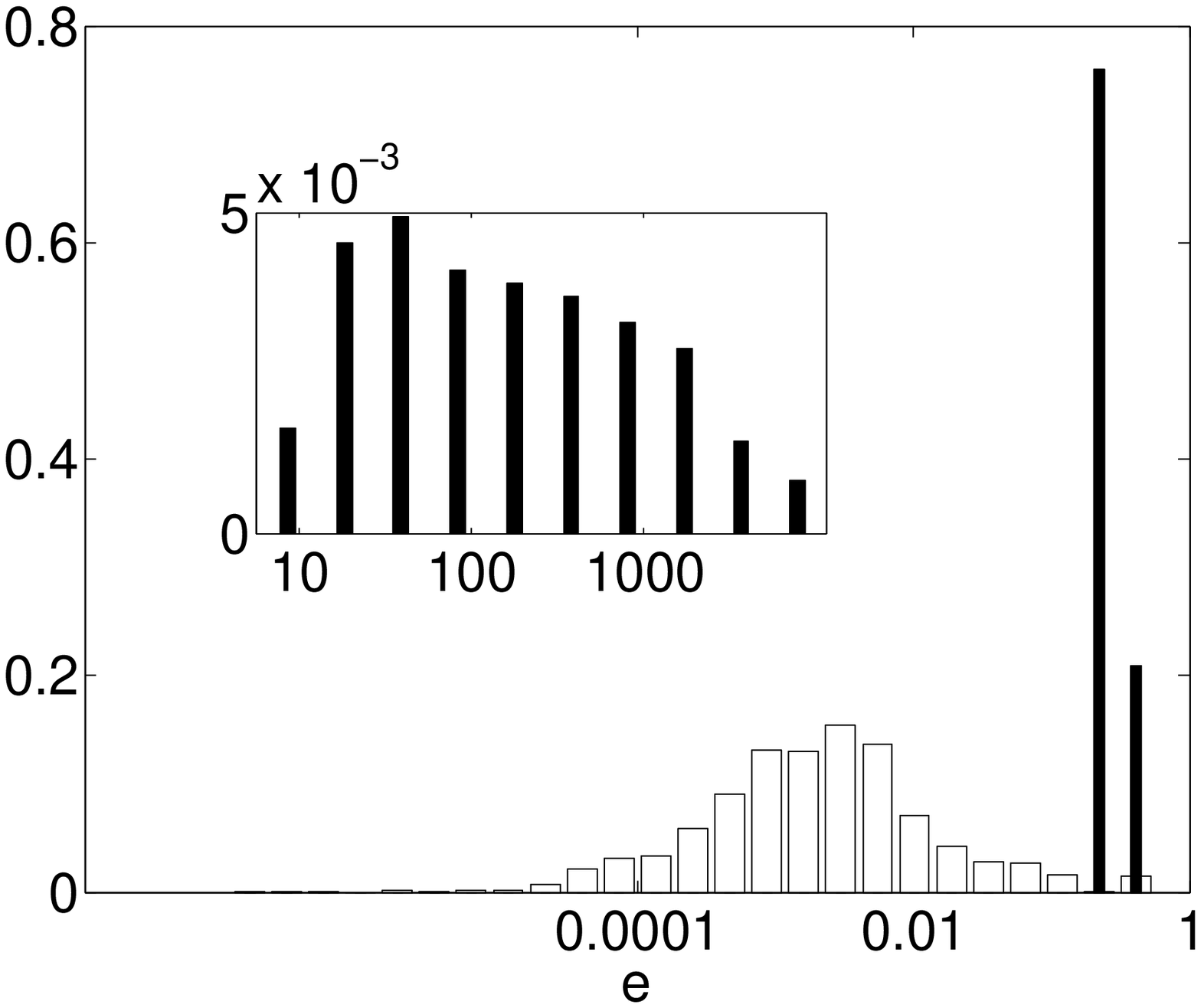}\\
    C) \includegraphics[width=0.3\textwidth]{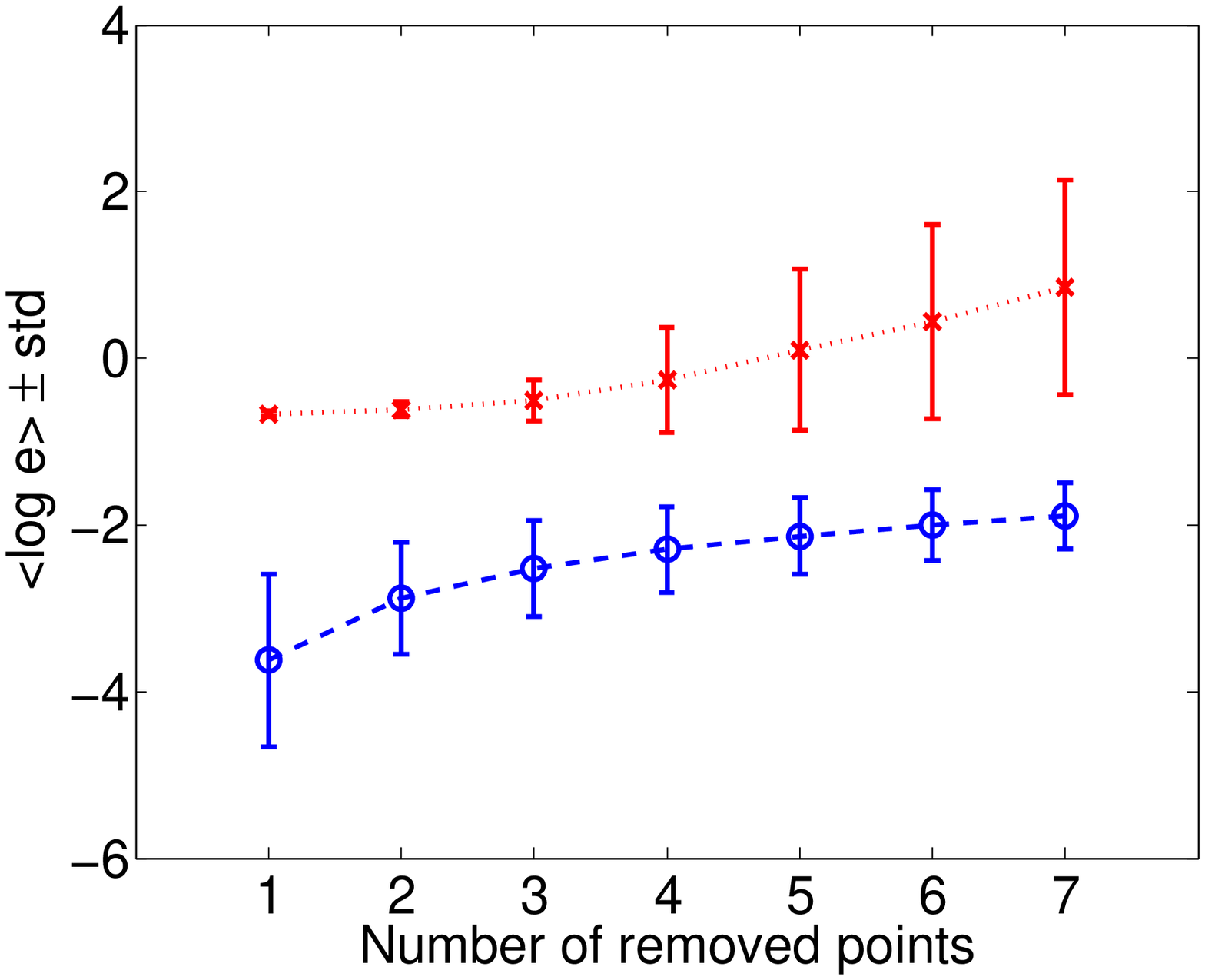}
    D) \includegraphics[width=0.3\textwidth]{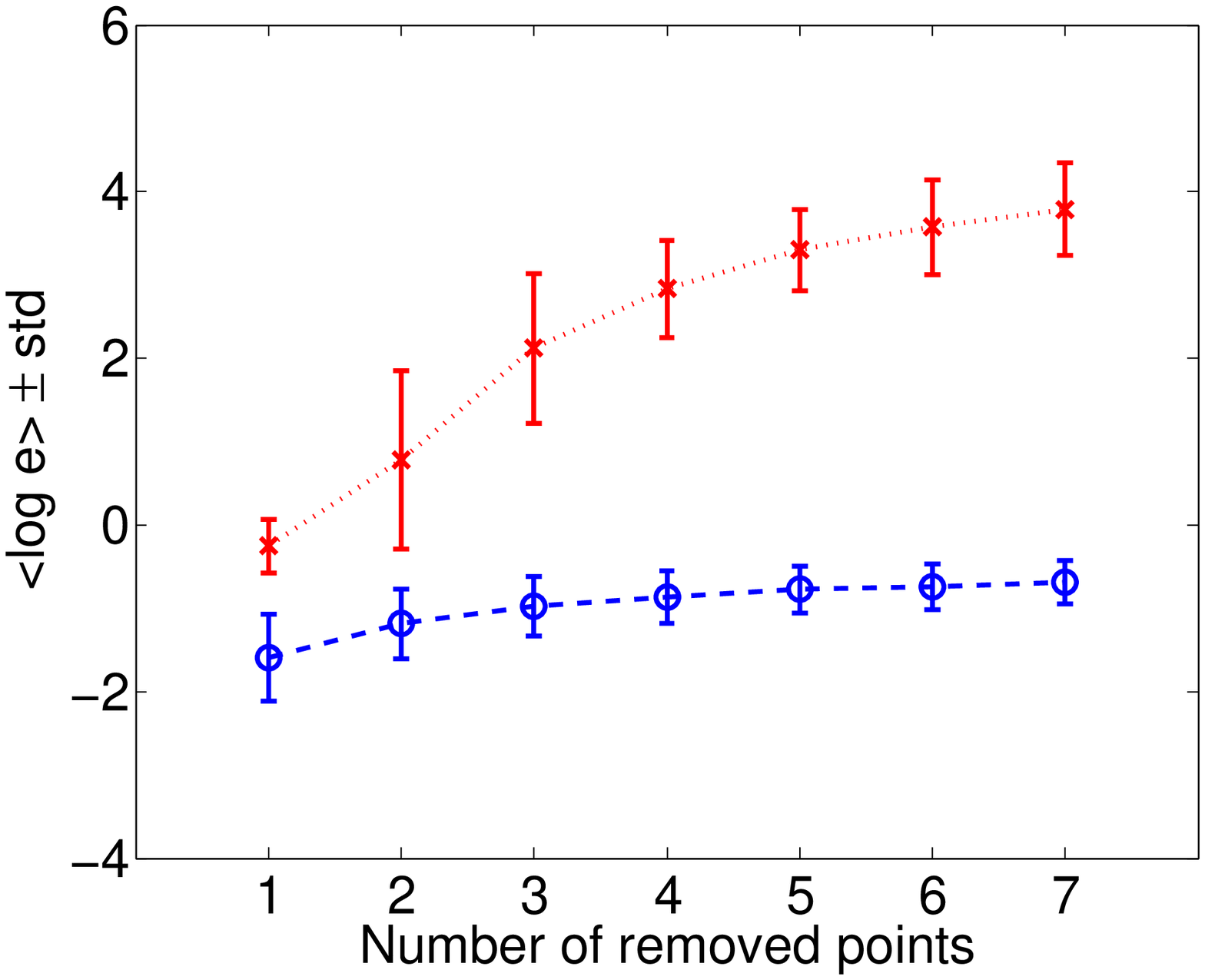}
  \end{center}
\caption{\label{figg2}Comparison of LA and LS methods of
  reconstructing CSD from incomplete data, see caption of
  Figure~\ref{figg}. The data used here are the same as in
  Figure~\ref{figg20}.}
\end{figure}

Clearly, performing reconstructions from incomplete data, we expect
the distortions to grow. Results of the test of the two methods are
shown in Fig.~\ref{figg2}. Fig.~\ref{figg2}A shows the results of the
iCSD reconstructions from data with one electrode removed. As in the
previous case, the distribution of errors of the LS method is
bimodal with very narrow modes, while the distribution of errors of
the LA method is rather broad (Fig.~\ref{figg2}B). However, unlike the
previous case, the LA method is almost always better.
This difference is preserved as the number of removed points is
increased. For both methods the mean error of reconstruction grows
with the number of removed points, which is expected. However, the
distribution of errors for the LS methods gets wider, while the
distribution of errors for LA method gets more narrow. This is true
for both the best 90\% cases and for the 10\% worst. In practice this means
that the LS method for such complicated CSD distributions is not
recommended.

\begin{figure}[htbp]
  \begin{center}
    \begin{tabular}[b]{c}
      A)\\[12.5em] B)
    \end{tabular} 
    \includegraphics[width=0.8\textwidth]{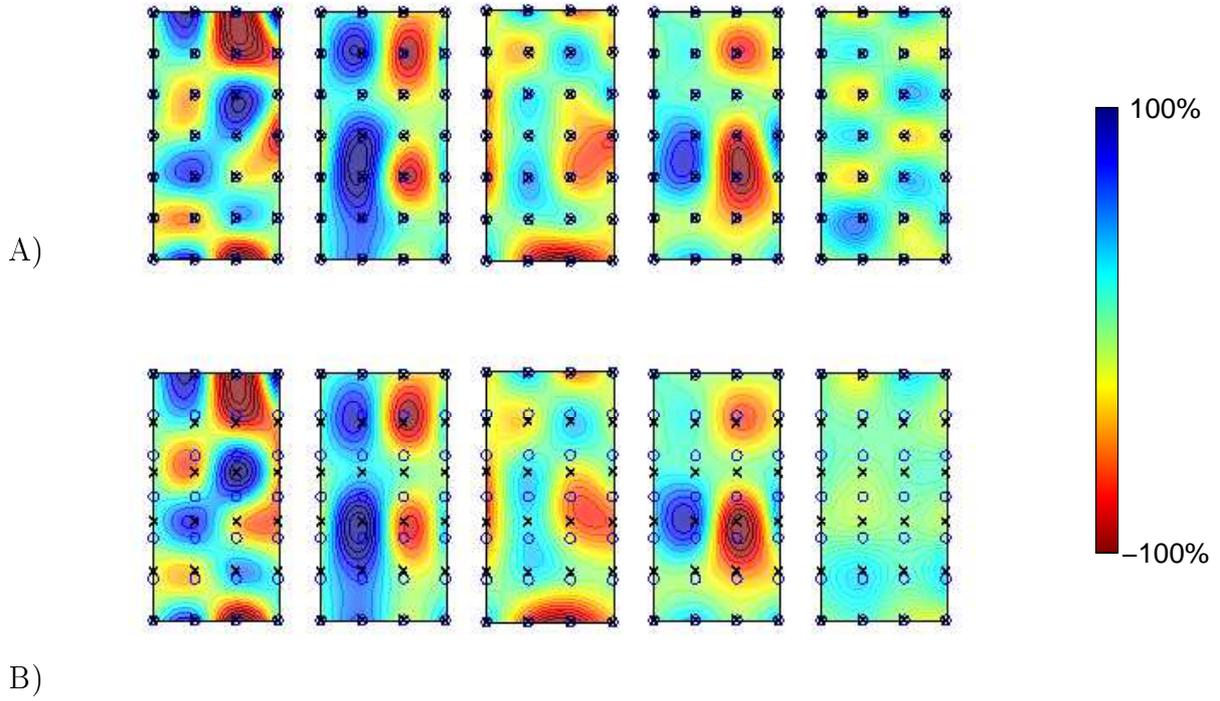} 
      \mbox{ } \raisebox{4em}[0pt]{
\includegraphics[width=0.1\textwidth]{colorbar.eps}}
  \end{center}
\caption{\label{figg30}Reconstructions of CSD from experimental data,
  $t=15$ms after stimulation of vibrissa. For description see caption
  of Fig.~\ref{figg20}.}
\end{figure}

\begin{figure}[htbp]
  \begin{center}
    A) \includegraphics[width=0.3\textwidth]{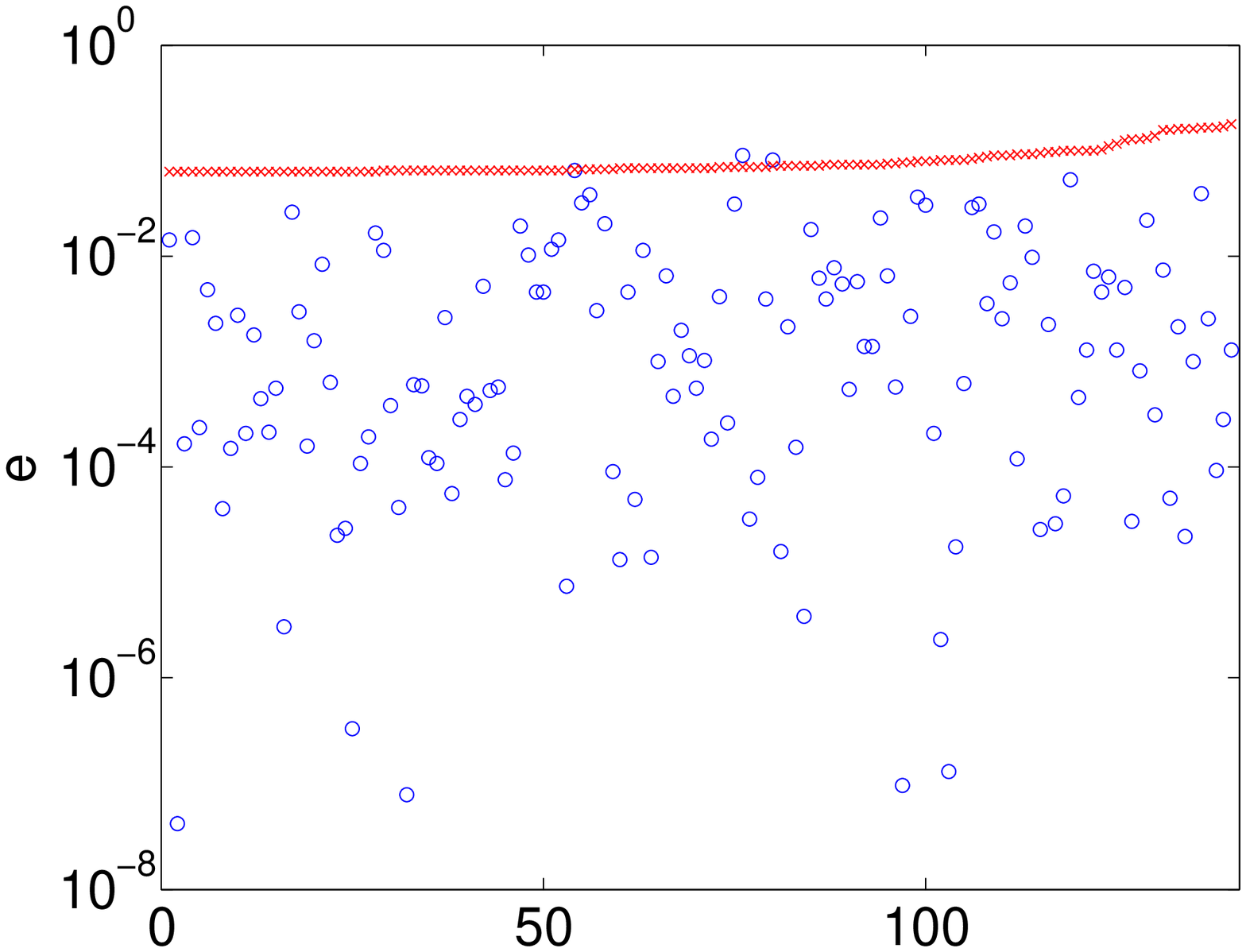}
    B) \includegraphics[width=0.3\textwidth]{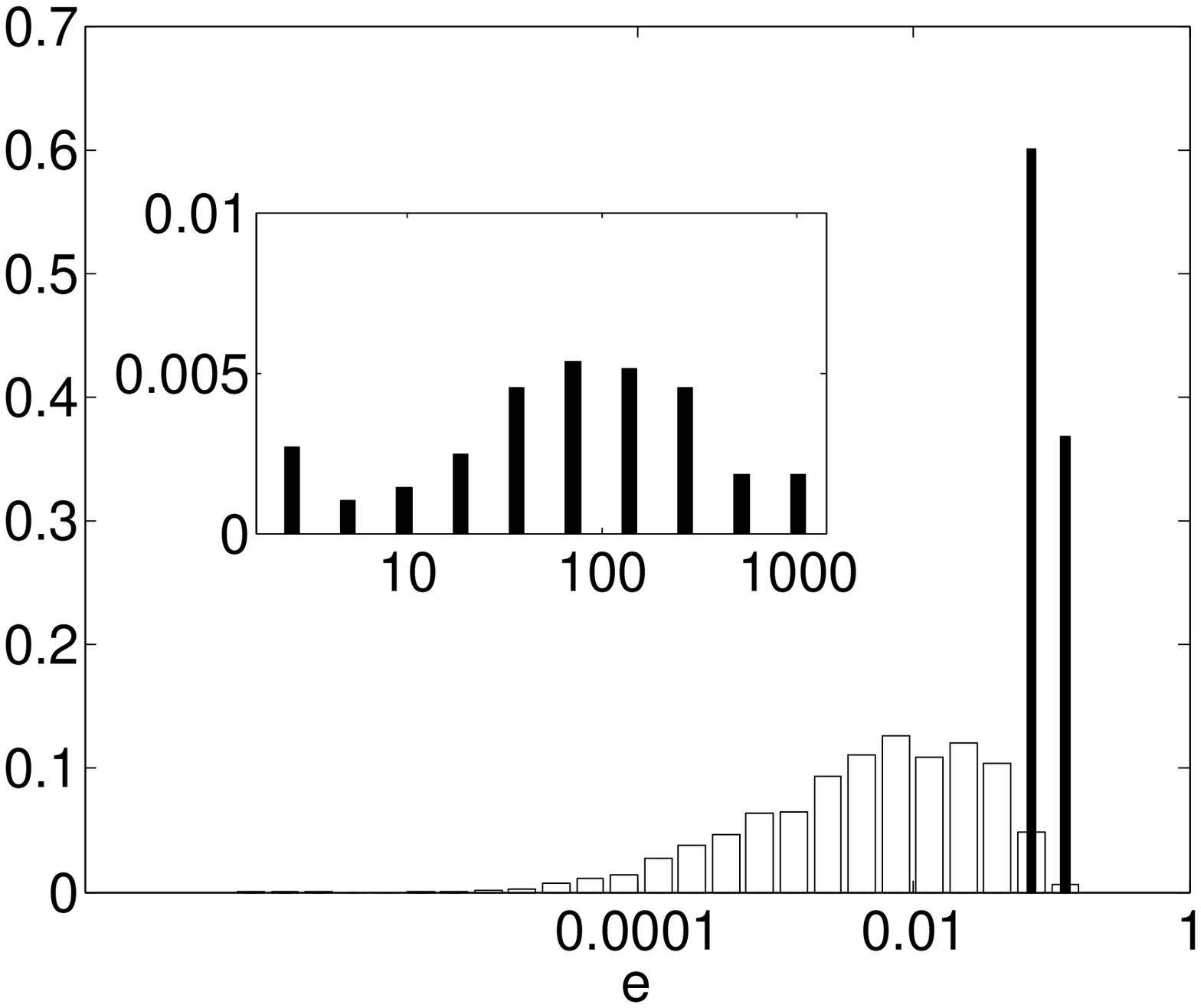}\\
    C) \includegraphics[width=0.3\textwidth]{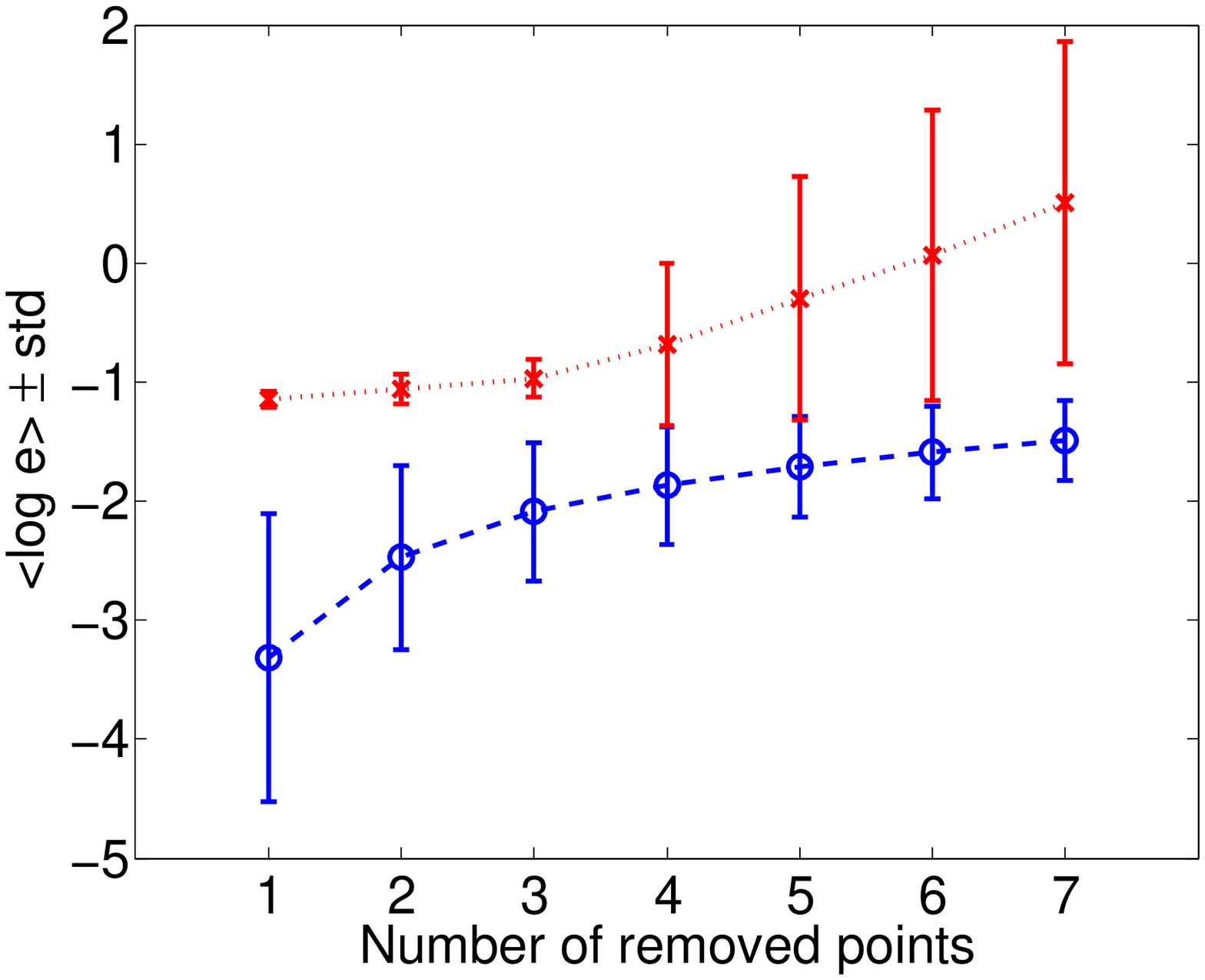}
    D) \includegraphics[width=0.3\textwidth]{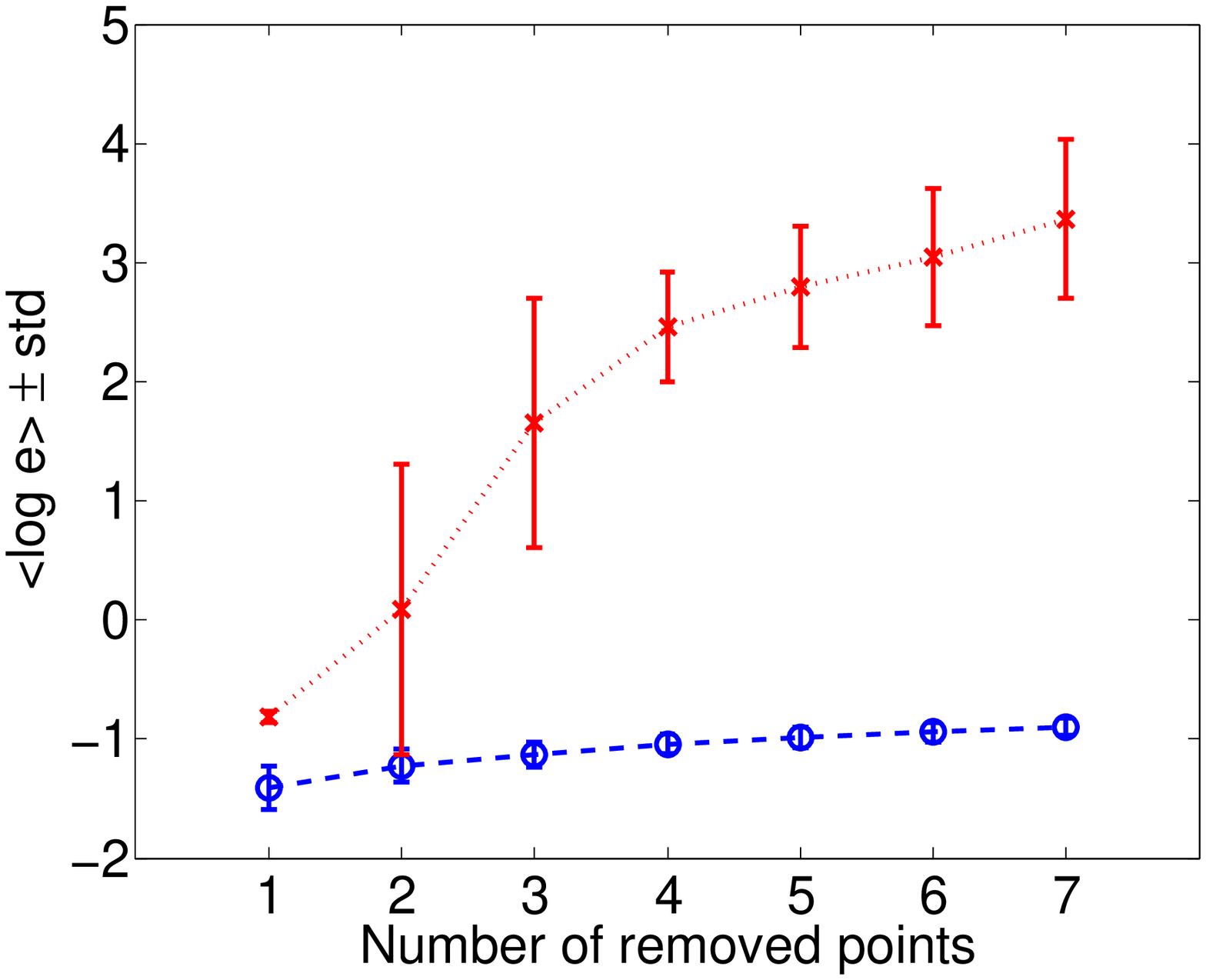}
  \end{center}

\caption{\label{figg3}Comparison of LA and LS methods of
  reconstructing CSD from incomplete data, see caption of
  Figure~\ref{figg}. The data used here are the same as in
  Figure~\ref{figg30}.}
\end{figure}

Such behavior was typical for this data set for the time frames we
inspected. For illustration and comparison we show the reconstructions
from the complete data on the original (Fig.~\ref{figg30}A) and
smaller (Fig.~\ref{figg30}B) grids
as well as the results of the same analysis for the recordings taken
15ms after the stimulus onset (Fig.~\ref{figg3}).

\section{Discussion}
\label{sec:discussion}

Reconstruction of the current source density generating recorded
extracellular potentials from these potentials is an ill-defined
problem.
The reason is that there is an infinite number of different
distributions which could lead to the same recordings. Nevertheless,
as the CSD is much more local reflection of the neural activity than
the potentials, there were many attempts to find a viable
reconstruction of the sources from the measured
fields~\citep{Nicholson1971, Mitzdorf1985, Pettersen2006}. One recent
candidate which has a number of advantages over the classical approach
is the inverse CSD method~\citep{Pettersen2006, Leski2007}. It has
been originally developed for situations where a set of recordings was
collected on a regular rectangular grid. Given the construction of the
method it is unclear how to proceed when one of the recordings is
missing, due to a failure of one of the electrode contacts or in other
cases. We have discussed here two approaches which might enable the
application of iCSD method to sets with incomplete data.

Local Averages method (LA) is simple, stable, and the results are
never very bad (normalized error of the order of a few percent), even for a
relatively large number of missing data points. The Least Squares
method (LS) seems attractive as it does not assume anything about the
missing data. The distribution of errors is usually bimodal with two
narrow modes. Usually, the errors are within a small range dominated
by the effect of the sparser grid, however, for a subset of cases 
which is growing with the number of missing data, the errors can be
extremely large.

The respective quality of reconstruction for the two methods depends
on the structure of the original sources, and (especially for LS) on the specific location
of the removed points. Our tests on the sources modeling the dipole
distributions of the cortex (Section~\ref{sec:results}) with the grid
shrinking along the dipole show that for a small number of missing
recording points (<5) the LS method usually gives smaller errors than
the LA. However, for more complex thalamic sources
(Section~\ref{sec:expdata}) the LA method is usually far better for
any number of removed points.

A priori it is not obvious which method to choose in analysis of
experimental data, when the original CSD is unknown and is to be
found. Our recommendation is to use the LA method in all
cases. Despite its simplicity it seems to be more stable and leads to
smaller errors, especially for complex distributions, thus it becomes
our method of choice. If the potential seems to vary relatively slowly
along one direction of the grid and the missing data are not nearest
neighbors lying at the edge, the LS method might also be worth
trying, but in general we do not recommend it.

One may wonder if it is possible to improve the technique beyond the
proposed approaches. One way would be to consider CSD distributions
spanned on the available recording points which would not necessarily
form a full regular grid. However, this seems rather difficult to
implement in full generality, as the spline coefficients would have to be
calculated from the scratch for every distribution of the recording
points, and the matrix connecting the potentials with the CSD
parameters would have to be calculated for every distribution adding
substantially to the computational overhead. A more promising approach
seems through the application of statistical 
methods. For example, one way we plan to follow in the future is to use an overdetermined basis of Gaussian sources
and search for efficient projections of the recordings on this
basis.

\appendix

\section*{Appendix: Gaussian test sources}
\label{sec:appGauss}

The Gaussian sources used in the test in Section~\ref{sec:results}
were of the form
\[
C(x,y,z) = 
\sum_{i=1}^8 A_i \exp
\left[
-\frac{(x-x_i)^2}{2(\sigma^{xz}_i)^2}
-\frac{(y-y_i)^2}{2(\sigma^{y}_i)^2}
-\frac{(z-z_i)^2}{2(\sigma^{xz}_i)^2}
\right],
\]
with the coefficients given in Table~\ref{tabela}. 
\begin{table}[htbp]
\begin{center}
\begin{tabular}{|l|c|c|c|c|c|c|c|c|}
\hline
 $i$ & 1 & 2 & 3 & 4 & 5 & 6 & 7 & 8 \\
\hline
 $x_i$ & 1 & 4 & 1 & 4 & 1 & 4 & 1 & 4 \\
 $y_i$ & 1 & 1 & 4 & 4 & 1 & 1 & 4 & 4 \\
 $z_i$ & 3.5 & 3.5 & 3.5 & 3.5 & 6.5 & 6.5 & 6.5 & 6.5 \\
 $\sigma^{xz}_i$ & 1 & 1 & 1 & 1 & 1 & 1 & 1 & 1 \\
 $\sigma^{y}_i$ & 1.5 & 1.5 & 1.5 & 1.5 & 1 & 1 & 1 & 1 \\
 $A_i$ & 0.8 & -1.1 & -1.2 & 1 & -1 & 1.2 & 0.5 & -0.9 \\
\hline
\end{tabular} 
\end{center}
\caption{\label{tabela}Coefficients of the Gaussian sources. Origin of the
  grid is $(x,y,z)=(1,1,1)$.  } 
\end{table}
Fig.~\ref{figg0} shows four parallel sections of the sources which
together pass through all the nodes of the grid (in the region spanned
by the virtual recording grid). To calculate the potentials we
truncated the sources to the region $-1\le x \le 6$, $-1\le y \le 12$,
$-1\le z \le 6$.

\section*{Acknowledgements}
\label{sec:ack}

This work was partly financed from the Polish Ministry of Science and Higher Education research grants N401 146 31/3239 and
PBZ/MNiSW/07/2006/11. SŁ was supported by the Foundation for Polish
Science.

\bibliographystyle{agsm}

\end{document}